\documentclass[superscriptaddress]{revtex4}
\usepackage{amsmath}
\usepackage{amsthm}
\usepackage{amsfonts}
\usepackage{graphicx}
\usepackage{bbold}
\usepackage{xcolor}

\begin{document}

\title{Quantum Artificial Life in an IBM Quantum Computer}

\date{\today}

\author{U. Alvarez-Rodriguez} 
\affiliation{Basque Centre for Climate Change (BC3), 48940, Leioa, Spain}
\affiliation{School of Mathematical Sciences, Queen Mary University of London, London E1 4NS, United Kingdom}
\affiliation{Department of Physical Chemistry, University of the Basque Country UPV/EHU, Apartado 644, 48080 Bilbao, Spain} 
\author{M. Sanz}
\affiliation{Department of Physical Chemistry, University of the Basque Country UPV/EHU, Apartado 644, 48080 Bilbao, Spain}
\author{L. Lamata}\email{lucas.lamata@gmail.com}
\affiliation{Department of Physical Chemistry, University of the Basque Country UPV/EHU, Apartado 644, 48080 Bilbao, Spain}
\author{E. Solano}
\affiliation{Department of Physical Chemistry, University of the Basque Country UPV/EHU, Apartado 644, 48080 Bilbao, Spain}
\affiliation{IKERBASQUE, Basque Foundation for Science, Maria Diaz de Haro 3, 48013, Bilbao, Spain}
\affiliation{Department of Physics, Shanghai University, 200444 Shanghai, China}

\begin{abstract} 
We present the first experimental realization of a quantum artificial life algorithm in a quantum computer. The quantum biomimetic protocol encodes tailored quantum behaviors belonging to living systems, namely, self-replication, mutation, interaction between individuals, and death, into the cloud quantum computer IBM {\it ibmqx4}. In this experiment, entanglement spreads throughout generations of individuals, where genuine quantum information features are inherited through genealogical networks. As a pioneering proof-of-principle, experimental data fits the ideal model with accuracy. Thereafter, these and other models of quantum artificial life, for which no classical device may predict its quantum supremacy evolution, can be further explored in novel generations of quantum computers. Quantum biomimetics, quantum machine learning, and quantum artificial intelligence will move forward hand in hand through more elaborate levels of quantum complexity.
\end{abstract}

\maketitle

\section{Introduction}
As described by Deutsch, a quantum computer is a device that intends to fulfill the Deutsch-Church-Turing principle, namely, to efficiently simulate a finitely realizable physical system in the framework of quantum mechanics~\cite{qc}. In this context, quantum supremacy would be reached when a quantum processor outperforms classical computers realizing a given task. Along these lines, several proposals to achieve quantum supremacy for a variety of quantum algorithms and quantum simulations have been proposed~\cite{qs1,qs2,qs3,qs4,qs5,qs6}. 

The keyword ``quantum" has overflowed the limits to which was initially constrained and, currently, incessantly spreads through the interdisciplinary scientific literature. Indeed, it is a source of inspiration for the breeding extensions of already existing models with their quantum counterparts~\cite{qh1,BennettBrassard,EkertCrypto,qh2,qh2Bis,qh3,qh3Bis,qh3Bis2,qh3Bis3,qh3Bis4}. Besides the appealing intellectual exercise, this hybridization is often motivated by a plausible improvement in the conditions and enhancement in the efficiency of the developed protocols. From all possible ramifications, including quantum machine learning and quantum artificial intelligence, our research in quantum biomimetics is concerned with the design of a framework for quantum algorithms based on the imitation of biological processes, belonging to the macroscopic classical complexity, and brought down by design to the microscopic quantum realm~\cite{Schrodinger,qbiom1,qbiom2,qbiom3,qbiom3Bis,qbiom4,qmem1,qmem2,qmem2Bis}. There may not always be a neat analogy between the physical models underlying our protocols and those used to describe real biological systems, but our proposed effective dynamics only partially aims at emulating core aspects of the mimicked process. From a wide perspective in the history of arts and science, close imitation is a natural first layer and wish in the aesthetic process. In this sense, plain simulation is a valid and fruitful engineering playground, where analogies abound and serve as communicating vessels between unconnected fields. Our central goal in quantum simulations and quantum computing is to go beyond it, through a higher creativity challenge, in the search of a second layer of a major art.
 
In the particular scenario of artificial life, simple models of organisms are able to undergo most common stages of life in a controlled virtual environment~\cite{al1,al2,al3}. When extending this to the quantum realm, particularities of quantum physics, such as its limitation to linear dynamics, the no-cloning theorem, or the exponentially growing dimensionality of Hilbert spaces, play a relevant role. The quantum artificial life protocol we have engineered and implemented goes beyond the straightforward quantization of existing classical models. In this sense, and with similar spirit of other contributions~\cite{qe1,qe2,qe3,qe4}, it is noteworthy to mention that we leave open the question whether the origin of life is genuinely quantum mechanical. What we prove here is that microscopic quantum systems can efficiently encode quantum features and biological behaviours, usually associated with living systems and natural selection.   

In this article, we report the first experimental implementation of a model for quantum artificial life~\cite{qbiom2} into a quantum computer. To this end, we make use of the facilities provided by the IBM {\it ibmqx4} quantum computing chip~\cite{ibm}. This work should be aligned with the ramping developments in classical and quantum machine learning and artificial intelligence: the development of algorithms and devices with the capacity to interpret and mimic human behaviors in order to solve useful problems and improve the interaction with human beings. Along these lines, we may foresee a future in which these idealized machines hybridize the knowledge in machine learning, artificial intelligence, and artificial life, with an internal structure and dynamics following the laws of quantum physics, as is already happening in the classical domain~\cite{f1}. 
\section{Results}

We begin with a brief description of the model for quantum artificial life \cite{qbiom2}, whose most important elements are the quantum living units or individuals. Each of them is expressed in terms of two qubits that we call genotype and phenotype. The genotype contains the information describing the type of living unit, an information that is transmitted from generation to generation. The state of the phenotype is determined by two factors: the genetic information and the interaction between the individual and its environment. This state, together with the information it encodes, is degraded during the lifetime of the individual. 

The goal of the proposed model is to reproduce the characteristic processes of Darwinian evolution, adapted to the language of quantum algorithms and quantum computing. The self-replication mechanism is based on two partial quantum cloning events, an operation that entangles either the genotype or the phenotype with a blank state and copies a certain expectation value of the original qubit in both of the outcome qubits. In this set of experiments, the self-replication consists in duplicating the expectation value of $\sigma_z$ in the genotype, in a blank state that will be transformed in the genotype of the individual in the next generation~\cite{qbiom1}. The process is completed by copying again $\sigma_z$ of the new genotype in another blank state that will be transformed in the phenotype of the new individual. The next subprotocol in the algorithm is the interaction between the individuals and the environment, which emulates the aging of living units until an asymptotic state that represents its death. This evolution is encoded in a dissipative dynamics that couples a bath with each of the phenotype qubits, with $\sigma=|0\rangle\langle 1|$ as Lindblad operator. The effective lifetime, i.e., the time the phenotype needs to arrive to the dark state of the Lindbladian up to a given error, depends implicitly on the genotype. The protocol also accounts for mutations, performed via random single qubit rotations in the genotype qubits or via errors in the self-replication process~\cite{qbiom2}. The final ingredient is the interaction between individuals, which conditionally exchange the phenotypes depending on the genotypes~\cite{qbiom2}. This behavior is achieved via a four-qubit unitary operation, where genotypes and phenotypes play the role of control and target qubits, respectively. The conjunction of these components leads to a minimal but consistent Darwinian quantum scenario. The protocol may be enriched when including spatial information, either quantum or classical, or increasing the model complexity by considering a larger set of observables.  

The first step for this implementation is to express each of the building blocks of the previous paragraph in terms of the quantum gates available in the superconducting circuit architecture of IBM cloud quantum computer~\cite{ibm}. Since we have selected $\sigma_z$ as the observable to clone, every partial quantum cloning event requires the realization of a $U_{\text{CNOT}}$ gate, that can be directly performed in the experiment. Regarding the interaction with the environment, we have adapted our protocol, because the experimental device does not allow to realize a conditional projection of the quantum state to the $|0\rangle$ in the phenotypes. The alternative we propose is to implement the transition between the basis states as a sequence of small rotations in $\sigma_y$ for the phenotype qubits, $e^{-i \sigma_y \theta}$, with $\theta$ tuned according to the duration of each simulated time step. We have employed $u_2(\phi,\lambda)$ and $u_3(\theta,\phi,\lambda)$ available in the experimental platform, to implement the single qubit gates.
\begin{equation}
u_2(\phi,\lambda)=\frac{1}{\sqrt{2}}\left( \begin{array}{cc} 1& - e^{i \lambda} \\ e^{i \phi} & e^{i (\lambda + \phi)} \end{array} \right) \hspace{.5 cm}  u_3(\theta,\phi,\lambda)=\left( \begin{array}{cc} \cos\frac{\theta}{2} & -e^{i \lambda} \sin\frac{\theta}{2} \\  e^{i \phi} \sin\frac{\theta}{2} &  e^{i (\phi+\lambda)} \cos\frac{\theta}{2}  \end{array} \right) 
\end{equation}

The gate $u_3(\theta,0,0)$ acting on genotype qubits can be used for the mutation events. Ideally, and in order to emulate their randomness both in the phase $\theta$ and in the presence or absence of the event, we could design the experimental runs following a classical program. For making the procedure tractable, we could discretize the range of $\theta$ in $n$ values, and divide the total experimental runs in $n+1$ groups to account for each of the different possibilities. The weight, or number of runs for each group, would  depend on our selection for the mutation rate as well as on the random parameters obtained with the external program. However, constrained by the flexibility of the experimental device, we propose a less realistic but pragmatic procedure: assume that the mutations will only be of a specific $\theta$, and therefore eliminate a source of randomness and diversity in the protocol. The single-qubit gate accounting for the mutations will be $\sigma_x$. Regarding the randomness in the presence or absence of mutation events, we will have to adapt our algorithm to perform the mutations in groups of $1024$ experimental runs, and achieve the mutation rate accordingly. The last subprotocol, the inter-individual interactions, requires the implementation of the interaction gate $U_I$, whose effect is to exchange two pairs of quantum levels, while leaving the rest unaltered, as $U_I |xxyy\rangle=|xyyx\rangle$ and $U_I|xyyx\rangle = |xxyy\rangle$, for $\{x,y\} \in \{0,1\}$. The challenge is to decompose $U_I$ in terms of the gates offered by the experimental setup. Our solution is given by $U_I=S_{23} U_{12} (\mathbb{1} \otimes F)U_{12} S_{23}$, with $F=U_{43}C_{34}C_{24}U_{23}C^{\dagger}_{34}U_{43}U_{23}$. Here, the first and second subindices denote the control and target qubit respectively, $U$ is the controlled-not gate, $S$ is the SWAP gate and $C$ is the controlled square root of not gate. These can be rewritten in terms of the controlled-not gate as $S_{ij}=U_{ij} U_{ji} U_{ij}$ and $C_{12}=( T \otimes P u_3(-\pi/4,0,0) ) U_{12} (\mathbb{1} \otimes u_3 (\pi/4, 0 ,0)) U_{12} (\mathbb{1} \otimes P^{\dagger})$, with $P=\sqrt{\sigma_z}$ and $T=\sqrt{P}$. An additional relation to point out is that the control target behavior in the controlled-not gate can be exchanged by introducing Hadamard gates, $U_{21}=(H\otimes H) U_{12} (H \otimes H)$. This is a useful formula for designing the quantum circuit in an experimental platform that only allows a single direction for the implementation of the $U_{\text{CNOT}}$.

\subsection{Experiments}
\paragraph*{Interaction between two individuals.}
We start with a quantum circuit designed for reproducing the dynamics of two interacting individuals. Two precursor genotypes are initialized in $|\psi\rangle_{g_1} = \cos \frac{\pi}{8} |0\rangle + \sin \frac{\pi}{8} |1\rangle$ and $|\psi\rangle_{g_2} = \cos \frac{3 \pi}{8} |0\rangle + \sin \frac{3 \pi}{8} |1\rangle$ with $u_3$. Afterwards, both individuals are completed by copying the genotype qubits in blank states via $U_{\text{CNOT}}$ gate, $|\psi\rangle_1 = \cos \frac{\pi}{8} |00\rangle + \sin \frac{\pi}{8} |11\rangle$ and $|\psi\rangle_2 = \cos \frac{3 \pi}{8} |00\rangle + \sin \frac{3 \pi}{8} |11\rangle$. In terms of $\theta_1=\pi/8$ and $\theta_2=3\pi/8$, the complete state, $|\psi_1\rangle \otimes |\psi_2\rangle$, reads
\begin{align*}
|\psi\rangle = \cos \theta_1 \cos \theta_2 |0000\rangle + \cos \theta_1 \sin \theta_2 |0011\rangle + \sin \theta_1 \cos \theta_2 |1100\rangle + \sin \theta_1 \sin \theta_2 |1111\rangle. 
\end{align*}
We now apply the interaction gate $U_I$ to conclude this building block,
\begin{align*}
U_I |\psi\rangle = \cos \theta_1 \cos \theta_2 |0000\rangle + \cos \theta_1 \sin \theta_2 |0110\rangle + \sin \theta_1 \cos \theta_2 |1001\rangle + \sin \theta_1 \sin \theta_2 |1111\rangle. 
\end{align*}
Notice that the interaction fully exchanges the phenotypes, $\langle \sigma_z \rangle_2$ and $\langle \sigma_z \rangle_4$, that are now equal to the opposite genotype, $\langle \sigma_z \rangle_1 = \cos^2 \theta_1 -\sin^2 \theta_1=\langle \sigma_z \rangle_4$ and $\langle \sigma_z \rangle_3 = \cos^2 \theta_2 -\sin^2 \theta_2=\langle \sigma_z \rangle_2$.

\begin{figure}[h!]
\begin{center}
\includegraphics[width=\textwidth]{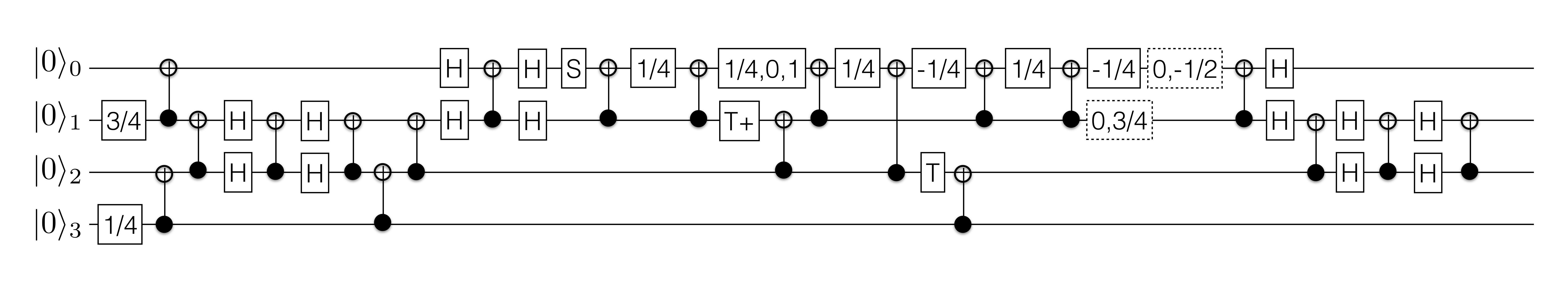}
\caption{Quantum circuit diagram for the protocol of two interacting individuals. Squares with a continuous line denote the phase values of $u_3$ gate, while squares with a dashed line denote the phase values of $u_2$ gate, both in units of $\pi$. When possible we reduce the expression to the value of $\theta$ and avoid writing the additional phases.}
\label{qcd1}
\end{center}
\end{figure}

The experiment is planned to reduce the total errors induced by the use of two-qubit gates. Consequently, we have reordered the initial Hilbert space $|g_1 p_1 g_2 p_2\rangle$, where $g_i$ is genotype and $p_i$ is phenotype, as $| p_2 g_2 p_1 g_1 \rangle$ and assigned each of these qubits to the experimental ones $|Q_0 Q_1 Q_2 Q_3 \rangle$. See Fig. \ref{qcd1} for the remaining quantum circuit diagram.

The results, in Table \ref{exp1}, agree with the ideal case with a $71.58 \%$ fidelity
according to $F(p,q)=\sum_j \sqrt{p_j q_j}$, that compares the probability distribution obtained when measuring in the computational basis with the theoretical prediction. Therefore, this result is valid, but not equivalent to the one that is expected when the complete wave function is considered, which is hindered by the use of full tomography and computed via $F(\rho_1,\rho_2)=\text{Tr} \sqrt{\sqrt{\rho_1} \rho_2 \sqrt{\rho_1}}$~\cite{r1}. The expectation values extracted from the data show a reasonable overlap between $p_1$ and $g_2$, as expected, and a considerable distance between $g_1$ and $p_2$.

\begin{table}[h!] 
\centering
\begin{tabular}{*{17}{|c}|}
\hline
Basis Element & 0000 & 0001 & 0010 & 0011 & 0100 & 0101 & 0110 & 0111 & 1000 & 1001 & 1010 & 1011 & 1100 & 1101 & 1110 & 1111 \\
\hline
Measured events & 1104 & 338 & 647 & 542 & 693 & 355 & 2687 & 519 & 104 & 144 & 114 & 1 & 99 & 132 & 261& 353 \\
\hline
Predicted events & 1012 & 0 & 0 & 0 & 0 & 0 & 5896 & 0 & 0 & 174 & 0 & 0 & 0 & 0 & 0 & 1012 \\
\hline
\end{tabular}
\caption{{\bf Interaction between two individuals.} Number of measurements for every element of the four-qubit basis. The experimental values for $\langle \sigma_z \rangle$ are (0.70,-0.26,-0.27,0.41) while the ideal values are (0.71,-0.71,-0.71,0.71). Notice that the mapping of reordering the qubits has been inverted to achieve the results in the $|g_1 p_1 g_2 p_2\rangle$ basis.}
\label{exp1}
\end{table}

\begin{figure}[h!]
\begin{center}
\includegraphics[width=0.5\textwidth]{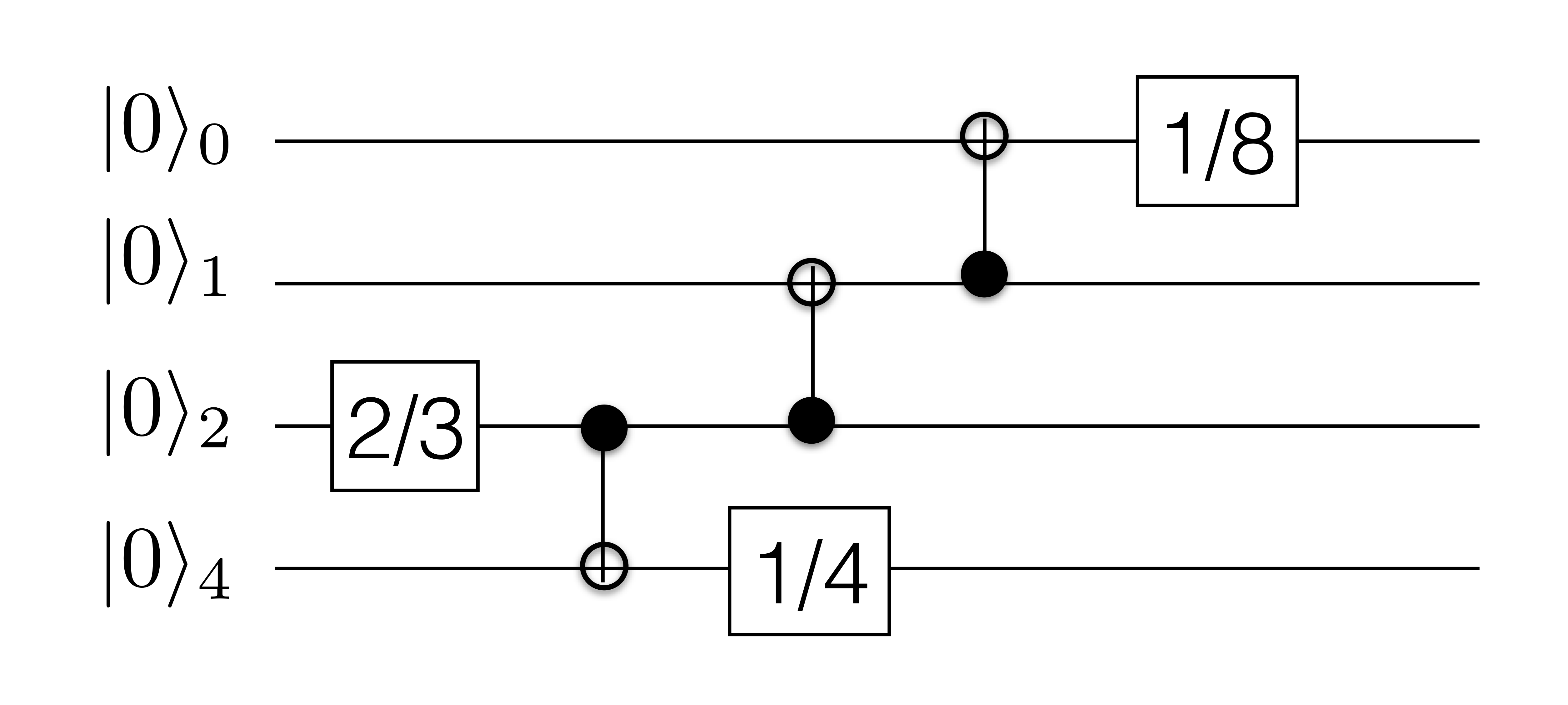}
\caption{The initialization of a genotype before three partial quantum cloning events. The first of these will produce an initial individual and the remaining two will replicate it into a second one. The protocol continues with single-qubit gates that emulate the dissipation. The squares denote $u_3(\theta,\phi,\lambda)$ gates where the number indicates the value of $\theta$.}
\label{qcd2}
\end{center}
\end{figure}

\paragraph*{Interaction with the environment.}
In this round of experiments we test the combination of partial quantum cloning events and dissipation. A precursor genotype is initialized in $|\psi\rangle_{g_1} = \cos \frac{\pi}{3} |0\rangle + \sin \frac{\pi}{3} |1\rangle$, and the individual completed with a first partial quantum cloning event via $U_{\text{CNOT}}$ and a blank state, $|\psi\rangle_1 = \cos \frac{\pi}{3} |00\rangle + \sin \frac{\pi}{3} |11\rangle$. Then, a single qubit rotation, $u_3(\pi/8,0,0)$, is applied in the phenotype, that substitutes the dissipation in a discrete manner, losing its exponential character. The course of time is simulated by this gate, by implementing one of them for every simulated time step. Subsequently, a second individual is created in a complete self-replication event with two partial quantum cloning operations. To conclude, $u_3(\pi/8,0,0)$ is implemented again on both genotypes associated with a next time step.
We assign the Hilbert space of the simulating device as $|Q_0 Q_1 Q_2 Q_4\rangle \rightarrow |p_2 g_2 g_1 p_1\rangle$ to maximize the efficiency of the protocol. See Fig. \ref{qcd2} for the quantum circuit diagram. The results, shown in Table \ref{exp2}, account for similar probability distributions between the ideal and the real data with a fidelity of $91.18\%$, as before computed only for the computational basis. 

\begin{table}[h!] 
\centering
\begin{tabular}{*{17}{|c}|}
\hline
Basis element & 0000 & 0001 & 0010 & 0011 & 0100 & 0101 & 0110 & 0111 & 1000 & 1001 & 1010 & 1011 & 1100 & 1101 & 1110 & 1111 \\
\hline
Measured events & 1491 & 103 & 42 & 224 & 387 & 46 & 31 & 249 & 67 & 91 & 108 & 916 & 149 & 354 & 439 & 3495 \\
\hline
Predicted events & 1682 & 66 & 0 & 0 & 288 & 11 & 0 & 0 & 0 & 0 & 34 & 866 & 0 & 0 & 200 & 5045 \\
\hline
\end{tabular}
\caption{{\bf Self-replication and interaction with the environment in the $\sigma_z$ basis.} Number of measurements for every element of the four-qubit basis. The experimental values for $\langle \sigma_z \rangle$ are (-0.37,-0.26,-0.34,-0.34) while the ideal values are (-0.5,-0.35,-0.5,-0.46). Notice that the mapping of reordering the qubits has been inverted to achieve the results in the $|g_1 p_1 g_2 p_2\rangle$ basis.}
\label{exp2}
\end{table}

For the self-replication instance, there is an additional property of the model that only arises when measuring some purely quantum correlations of the system. The partial quantum cloning operation entangles the qubits which are involved on it, transmitting $\langle \sigma_x\rangle$ of the original state into $\langle \sigma_x \otimes \sigma_x \rangle$. Note that this data can be extracted from the experiment when measuring on the $\sigma_x$ basis, which is done by introducing a Hadamard gate in every entry before projecting. Therefore, one has to compute $\langle \sigma_z \otimes \sigma_z \otimes \sigma_z \otimes \sigma_z  \rangle$ in the new basis, to retrieve $\langle \sigma_x \otimes \sigma_x \otimes \sigma_x \otimes \sigma_x  \rangle$. This technique is based on the equality $\rm{Tr}[\sigma_x \rho]=\rm{Tr}[\sigma_z H \rho H]$, since $\sigma_x = H \sigma_z H$. Even if the calculation for the fidelity yields a satisfactory $93.45 \%$, the value of $\langle \sigma_x \otimes \sigma_x \otimes \sigma_x \otimes \sigma_x  \rangle$ still shows a sizable error with respect to the ideal one, as we show in Table \ref{exp3}.

\begin{table}[h!] 
\centering
\begin{tabular}{*{17}{|c}|}
\hline
Basis element & 0000 & 0001 & 0010 & 0011 & 0100 & 0101 & 0110 & 0111 & 1000 & 1001 & 1010 & 1011 & 1100 & 1101 & 1110 & 1111 \\
\hline
Measured events & 753 & 246 & 277 & 52 & 448 & 747 & 569 & 513 & 343 & 493 & 616 & 177 & 717 & 679 & 345 & 749 \\
\hline
Predicted events & 624 & 1 & 77 & 547 & 157 & 1150 & 704 & 603 & 77 & 547 & 624 & 1 & 704 & 603 & 157 & 1150 \\
\hline
\end{tabular}
\caption{{\bf Self-replication and interaction with the environment in the $\sigma_x$ basis.} Number of measurements for every element of the four-qubit basis rotated to $\sigma_x$. The experimental value for $\langle \sigma_x \otimes \sigma_x \otimes \sigma_x \otimes \sigma_x  \rangle$ is 0.22 while the ideal value is 0.56. Again, the qubits have been reordered to coincide with the ideal results in the $|g_1 p_1 g_2 p_2\rangle$ basis. An additional point to remark here is the fact that the global value of $\langle \sigma_x \otimes \sigma_x \otimes \sigma_x \otimes \sigma_x  \rangle$ is different to the product of $\langle \sigma_x \otimes \sigma_x \rangle$ for each individual, which yields a value of 0. This calculation shows that both individuals are indeed causally related.}
\label{exp3}
\end{table}

Even if this implementation does not coincide with the time evolution presented in the original model, it is able to emulate its results when only focusing on the $\sigma_z$ or $\sigma_x$ basis, but not to compare both measurements in general. Accordingly, if the lifetimes of each living qubits undergo a similar dynamics to the ones proposed in the model, the effect of the environment on the correlations cannot be correctly reproduced, and viceversa, unless the gates are specifically selected for a given precursor genotype. The theoretical value of $\langle \sigma_x \otimes \sigma_x \otimes\sigma_x \otimes \sigma_x \rangle$ for a system that only undergoes self-replication events and dissipation decreases as $\langle \sigma_x \rangle e^{-\gamma (t_1 + t_2)/2}$, with $\langle \sigma_x \rangle$ calculated over the precursor genotype and $t_i$ being the time between self-replication events.  For the variant of the single qubit gates analyzed here, the theoretical value goes as $\langle \sigma_x \rangle \cos \theta_1 \cos \theta_2$, where $\theta_i$ indicate the phase of each $u_3 (\theta_i,0,0)$. This part of the dissipative dynamics should match with the evolution of $\langle \sigma_z \rangle$, so the following set of equations should be fulfilled:
\begin{eqnarray}
\label{lio}
e^{-\gamma(t_1 +t_2)/2} \langle \sigma_x \rangle &=& \cos \theta_1 \cos \theta_2 \langle \sigma_x \rangle \\
\nonumber 1- 2 e^{-\gamma (t_1 + t_2)}(1-a) &=& (2a -1)\cos \theta_1 \\
\nonumber 1- 2 e^{-\gamma t_2}(1-a) &=& (2a -1) \cos \theta_2 
\end{eqnarray} 
where $a$ is the $|0\rangle\langle0|$ component in the precursor genotype. Given that there is no solution for $\theta_1$ and $\theta_2$ which is independent of $a$, the method of single-qubit gates for mimicking the dissipation is not valid as a general protocol, because it has to be tuned for each case. Nevertheless, the important quantum feature of the model, the existence of quantum correlations, and their role as witnesses of the interactions between quantum living units can correctly be represented with the approach followed here, even if their time dependence is different to the one presented in the original model. 

In more practical terms, the implementations summarized in Table II and Table III are realistic, but not compatible between them, because both can be associated to dissipative dynamics but with different representative parameters, as we have seen in Eq. \eqref{lio}. Furthermore, we believe that the ideal realization of the experiment will soon be feasible at least for a small number of individuals. Our proposal for introducing the dissipation is to exploit the natural decoherence present in quantum platforms, and use error correction protocols only in the genotype qubits. This phenotype-genotype asymmetry in the decay probability is the key element in the emulation of the interaction between individuals and environment. 

\begin{figure}[h!]
\begin{center}
\includegraphics[width=\textwidth]{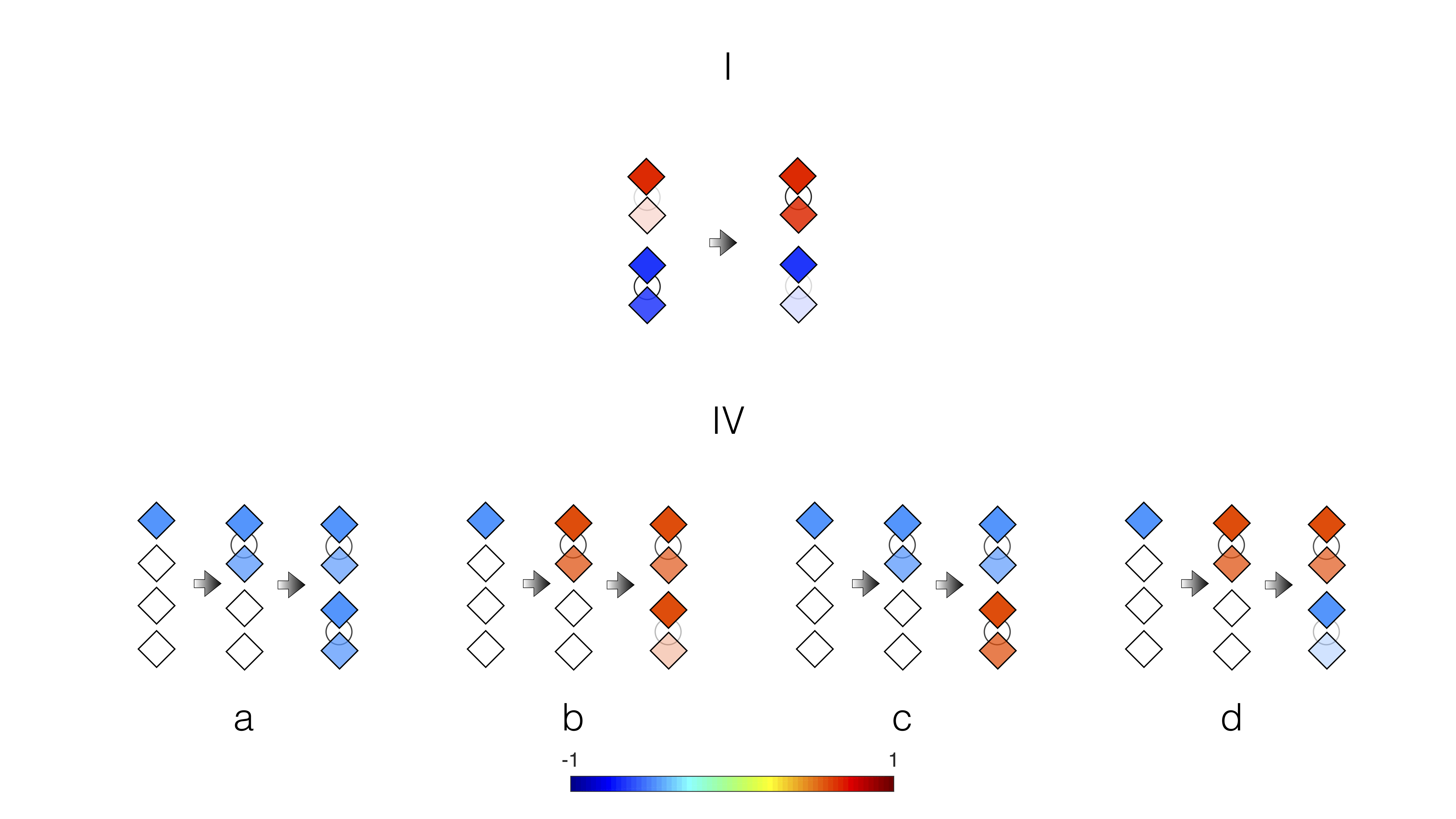}
\caption{Visualization of the ideal processes in experiments I and IV. We depict the individuals as combinations of two diamonds that represent the genotype and phenotype qubits. The color in the genotype qubit, the upper diamond of each pair, depends on the value of $\sigma_z$ as indicated in the color bar. The color in the phenotype qubit is the same as in the genotype one, as the color is meant to be showing the genetic information. Moreover, the opacity of this color is modified according to the expectation value of $\sigma_z$ being limited by the value of $1$ that corresponds to the blank qubits. In both cases the right arrow separates two consecutive time steps. Following these clarifications, we can see the exchange of phenotypes in I, and the self-replication followed by different mutation possibilities in IV.}
\label{vis}
\end{center}
\end{figure}

The implementation of mutations requires to combine the outcome of different designs of quantum circuit diagrams and, therefore, experimental runs. In this case, we consider that a mutation event, which can affect both individuals, is simulated with a $\sigma_x$. The complete result is achieved when gathering data from $4$ different groups of experiments, that correspond to the cases of mutation on the first genotype, mutation on the second genotype, mutation on both genotypes, and no mutation. We have performed $1024$ experimental runs for each of the three cases with mutations and $8192$ runs for the no-mutation rate. These results have been combined with the ones shown in Table \ref{exp2}, that coincide with the no-mutation case, with the goal of reducing the mutation rate for each individual, which takes a final value of $2/19$. Table \ref{exp4} contains the agreggated data of the mutation experiments.  In IVb the mutation occurs before the self-replicating event, therefore affecting the second individual, in IVc the mutation can only occur after the second individual has been created, and IVd contains both mutations. See the illustration of this process in Fig. \ref{vis}. See Table \ref{exp4} for the measured data with a fidelity of $94.86\%$ with respect to the ideal case in the $\sigma_z$ basis.

\begin{table}[h!] 
\centering
\begin{tabular}{*{17}{|c}|}
\hline
Basis element & 0000 & 0001 & 0010 & 0011 & 0100 & 0101 & 0110 & 0111 & 1000 & 1001 & 1010 & 1011 & 1100 & 1101 & 1110 & 1111 \\
\hline
Measured events & 3201 & 282 & 385 & 447 & 1344 & 217 & 541 & 494 & 235 & 608 & 297 & 2026 & 464 & 1509 & 946 & 6325 \\
\hline
Predicted events & 3466 & 137 & 309 & 12 & 1174 & 46 & 622 & 25 & 12 & 303 & 76 & 1930 & 26 & 659 & 401 & 10123 \\
\hline
\end{tabular}
\caption{{\bf Self-replication, interaction with the environment and mutations.} Number of measurements for every element of the four-qubit $\sigma_z$ basis. The experimental values for $\langle \sigma_z \rangle$ are (-0.28,-0.23,-0.19,-0.23) while the ideal values are (-0.40,-0.35,-0.40,-0.37). The experimental basis is also permuted to coincide with the ideal results in the $|g_1 p_1 g_2 p_2\rangle$ basis.}
\label{exp4}
\end{table}

\begin{figure}[h!]
\begin{center}
\includegraphics[width=\textwidth]{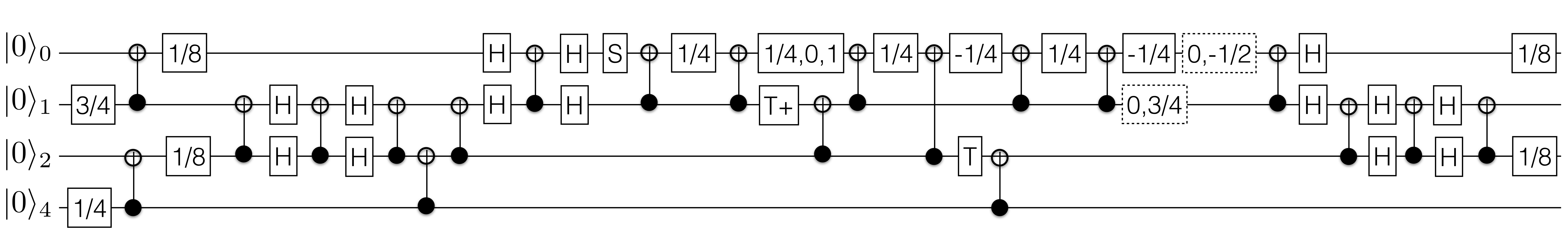}
\caption{Quantum circuit diagram for the complete quantum artificial life protocol. Squares with a continuous line denote the phase values of $u_3$ gate, while squares with a dashed line denote the phase values of $u_2$ gate, both in units of $\pi$. When possible we reduce the expression to the value of $\theta$ and avoid writing the additional phases.}
\label{qcd3}
\end{center}
\end{figure}

\paragraph*{Realization of the complete model of quantum artificial life.}
The last round of experiments is devoted to the reproduction of the aggregate of properties in the quantum artificial life algorithm. In order to maintain the fidelity in values that allow us to claim that the experiment is indeed behaving according to the protocol, we restrict our analysis to the case of two interacting individuals, which undergo mutations and dissipation. Then, the quantum circuit diagram, shown in Fig. \ref{qcd3}, is an upgraded version of the one shown in Fig. \ref{qcd1} that includes $u_3(\pi/8,0,0)$ for simulating the dissipation in the phenotypes. For the mutations, we follow the same strategy as in the previous subsection, combining the data generated with different quantum circuit diagrams each of them emulating a specific case of the presence or absence of mutation instances. In particular, $3$ rounds of $8192$ runs emulating the no-mutation case and $1024$ runs for each of the mutation cases determine a mutation rate of $2/27$. The post-processing of the data, in Table \ref{exp5}, matches the ideal probability distribution in the computational basis with a fidelity of $93.94\%$. 

\begin{table}[h!] 
\centering
\begin{tabular}{*{17}{|c}|}
\hline
Basis element & 0000 & 0001 & 0010 & 0011 & 0100 & 0101 & 0110 & 0111 & 1000 & 1001 & 1010 & 1011 & 1100 & 1101 & 1110 & 1111 \\
\hline
Measured events & 3449 & 1598 & 2656 & 2053 & 2361 & 298 & 6369 & 2166 & 521 & 500 & 629 & 529 & 594 & 692 & 656 & 1146 \\
\hline
Predicted events & 2221 & 924 & 2221 & 410 & 410 & 251 & 12401 & 2207 & 237 & 410 & 237 & 838 & 584 & 251 & 410 & 2207 \\
\hline
\end{tabular}
\caption{{\bf Complete model.} Number of measurements for every element of the four-qubit basis. The experimental values for $\langle \sigma_z \rangle$ are (0.60,-0.09,-0.24,0.31) while the ideal values are (0.60,-0.43,-0.60,0.43). The experimental basis is also permuted to coincide with the ideal results in the $|g_1 p_1 g_2 p_2\rangle$ basis.}
\label{exp5}
\end{table}

\begin{figure}[h!]
\begin{center}
\includegraphics[width=\textwidth]{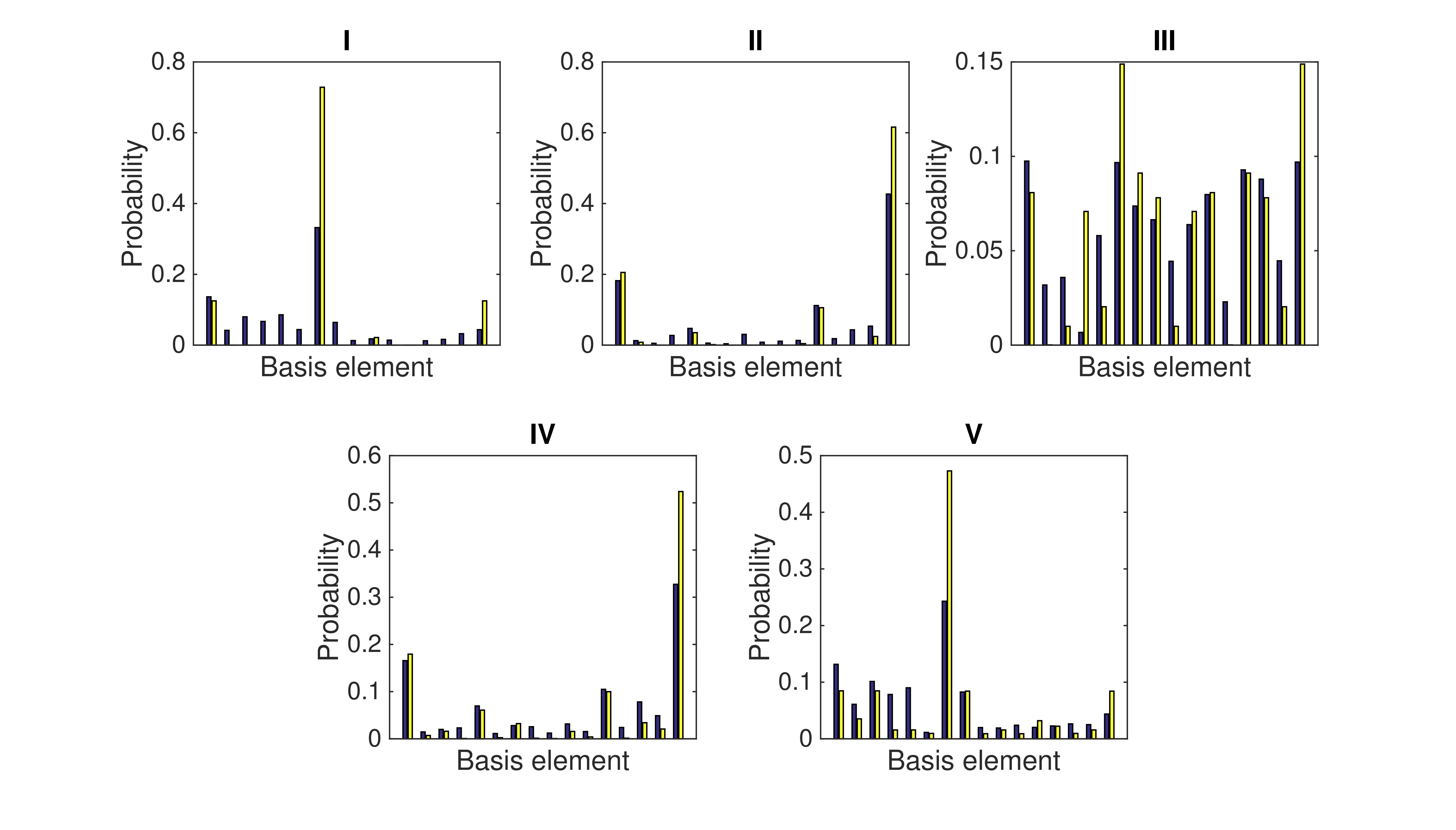}
\caption{Experimental and ideal probability distributions for all the cases analyzed. In each pair of columns the blue one in the left denotes the experimental value while the yellow one in the right denotes the ideal estimation. The labels in the subplots correspond to the tables for the different quantum artificial life instances.}
\label{prob}
\end{center}
\end{figure}

\section{Discussion}

\begin{table}[h!] 
\centering
\begin{tabular}{*{17}{|c}|}
\hline
 & 0000 & 0001 & 0010 & 0011 & 0100 & 0101 & 0110 & 0111 & 1000 & 1001 & 1010 & 1011 & 1100 & 1101 & 1110 & 1111 \\
\hline
I & 1104 & 338 & 647 & 542 & 693 & 355 & 2687 & 519 & 104 & 144 & 114 & 1 & 99 & 132 & 261& 353 \\
\hline
II & 1491 & 103 & 42 & 224 & 387 & 46 & 31 & 249 & 67 & 91 & 108 & 916 & 149 & 354 & 439 & 3495 \\
\hline
III & 753 & 246 & 277 & 52 & 448 & 747 & 569 & 513 & 343 & 493  & 616 & 177 & 717 & 679 & 345 & 749 \\
\hline
IVa & 1511 & 121 & 46 & 201 & 395 & 86 & 27 & 213 & 102 & 192 & 162 & 984 & 215 & 635 & 478 & 2824 \\
\hline
IVb & 136 & 14 & 3 & 14 & 542 & 32 & 9 & 6 & 12 & 29 & 16 & 122 & 17 & 12 & 1 & 5 \\
\hline
IVc & 39 & 28 & 149 & 7 & 13 & 34 & 46 & 3 & 22 & 137 & 6 & 2 & 68 & 444 & 16 & 1 \\
\hline
IVd & 24 & 16 & 145 & 1 & 7 & 19 & 428 & 23 & 32 & 159 & 5 & 2 & 15 & 64 & 12 & 0 \\
\hline
Va & 1032 & 422 & 756 & 594 & 737 & 51 & 2121 & 731 & 88 & 109 & 132 & 113 & 99 & 179 & 186 & 383 \\
\hline
Vb & 1087 & 414 & 863 & 696 & 697 & 44 & 2076 & 692 & 107 & 107 & 134 & 97 & 99 & 191 & 188 & 304 \\
\hline
Vc & 1046 & 445 & 871 & 631 & 723 & 46 & 2056 & 634 & 111 & 114 & 154 & 148 & 97 & 162 & 192 & 348 \\
\hline
Vd & 7 & 6 & 48 & 45 & 47 & 5 & 24 & 61 & 99 & 89 & 119 & 122 & 39 & 48 & 35 & 68 \\
\hline
Ve & 219 & 269 & 62 & 63 & 74 & 87 & 44 & 27 & 27 & 15 & 15 & 7 & 28 & 57 & 12 & 18 \\
\hline
Vf & 58 & 42 & 56 & 24 & 83 & 65 & 48 & 21 & 89 & 66 & 75 & 42 & 232 & 55 & 43 & 25 \\
\hline
\end{tabular}
\caption{{\bf Experimental data.} The rows denote the number of events in each of the elements of the computational basis for all the instances considered. For the composite experiments, in order to retrieve the measurements shown in previous tables, one has to add all the events in each of the individual runs. In the particular case of IV, the data in II also contributes to the final result.}
\label{exp}
\end{table}

\subsection{Quantum vs Classical}
A natural method to evaluate the Quantum Artificial Life framework is to clearly describe on the similarities and differences between the quantum model and a classical analogue approach. On the one hand, all the indicators based in measurements on the $\langle \sigma_z \rangle$ basis can be reproduced by classical probability distributions. This means, that one can create a classical model of interacting individuals with identical ingredients, in which the single qubit results for each of the living units would be equal to the ones achieved in the quantum version. On the other hand, non-zero quantum correlations in particular, $\langle \sigma_x \otimes \sigma_x \rangle$, and its generalization to more pairs of qubits, can only be achieved in the quantum case. These introduce a new feature when compared with the classical version of the model, as they can be interpreted as a time correlations between the quantum living units. Let this be illustrated with the following example. Suppose we have two pairs of quantum living units, all of them with the same value of $\langle \sigma_z \rangle$ in the genotype qubit but with different values for the phenotype, in a simplified system without  mutations and interindividual interactions. When the individuals are not causally connected, the value of $\langle \sigma_x \otimes \sigma_x \otimes\sigma_x \otimes \sigma_x \rangle$ in the four qubits of both individuals would yield a value of $\alpha ^2$. However, if these individuals are related by a self-replication operation, the measurement would yield a value of $\alpha$, where $\alpha$ accounts for $\langle \sigma_x \rangle$ in the precursor genotype. In other words, the nonzero quantum correlation between the subspaces of the first and second individuals allows to discriminate two different operations from the timeline perspective.

One can show that a classical counterpart of this model would not be able to store the information of the time correlations. The correlations we are interested in are originated in the initial nonzero value of $\langle \sigma_x \rangle$ the precursor genotype, and afterwards propagated in each partial quantum cloning event because of the property mentioned above. In order to validate our claim, let us assume that the precursor genotype is given by an incoherent mixture of states of the form $\rho = a |0\rangle\langle0| + (1-a) |1\rangle\langle 1|$. Already from this point, one can see that $\langle \sigma_x \rangle = 0$, and therefore no information is going to be propagated to the global correlations when more than one qubit is present. In other words, the purely quantum information is produced by a quantum superposition state between the basis elements, and afterwards propagated through the entanglement of quantum living units. The interpretation of this phenomenon is that the quantum model allows to keep track of the relation between the individuals without the need of introducing additional variables to the ones that describe the genotype and phenotype. 

\subsection{Experimental Errors}

Regarding errors in the experimental protocol, even if the fidelities achieved are satisfactory, they do not correspond to the fidelities of the complete quantum state. In this sense, the prediction for the number of events to measure is done by simply multiplying the probability distribution by the number of events. These do not exactly match the experimental data (see Table \ref{exp} for all experimental outcomes). Indeed, the distances have to be properly weighed, since the probability distribution is the key quantity to be extracted (see Fig. \ref{prob}). Moreover, the overlap between expectation values of observables in the measurement basis is lower than the fidelity of the probability distribution for all cases analyzed here. 

In parallel, the assignment between the simulated and the simulating Hilbert spaces is designed to maximize the fidelity according to the calibration parameters provided by IBM. Nevertheless, the recalibration of the circuit changes the gate and readout errors, so we reevaluate our circuit according to the new parameters and adapt it when the fidelity outcome can be improved with a different labeling of qubits. Consequently, the performance of the different experiments is not directly comparable, since they have been implemented under unequal conditions. 

Despite the different factors degrading the implementation, the performed experiments reproduce the characteristic properties of the sought quantum natural selection scenario. We have observed how the partial quantum cloning events allow us to inherit the information of $\langle \sigma_z \rangle$ from qubit to qubit, and use this property to encode the self-replication process. We have also seen how nonzero quantum correlations assure that both individuals have been part of a same event in their timelines, in this case self-replication. Another relevant characteristic of the analysis is the inclusion of mutations as a source of randomness that, counterintuitively, significantly improved the fidelity of the quantum algorithm outcome. Our explanation is that mutations tend to homogenize the probability distribution, which is the same effect as the one produced by the errors naturally present in the experimental platform. Surprisingly, in the classical realm, mutations also help the species to adapt to changing environments.

\subsection{Scope of Quantum Artificial Life}

Regarding the emergence of complexity, the route towards the scalability of our quantum algorithm is intrinsically related to the inclusion of more degrees of freedom in the description of quantum living units. These may be introduced by simply increasing the number of qubits, and making them part of the updated genotype and phenotype. A part of the dynamics would be adapted by repeating the partial cloning processes and extending the dissipation to the new phenotype qubits. Another part of the dynamics would deal with the properties introduced by the new degrees of freedom. In the same way as the genotype in the current model rules the individual-environment and inter-individual interactions, additional observables in the genotype would enable the exploration of more characteristics: different self-replication rates, independent lifetime and interaction role, or capacity to displace along the associated Hilbert space, all of them encoded in the genotype. An alternative is to encode the information in quantum states of higher dimensions. The general result of partial quantum cloning to qudits of any dimension makes this family of hypothetical models feasible, conditional to the availability of high dimensional entangling operations.

A different question is the scalability of the current model without including any modification. Notice that, in our protocol, the information about the lifetime and the predator-prey character is classical and encoded in the mean value $\langle \sigma_z \rangle$ of the phenotype and the genotype of the individual, respectively. Therefore, this partial dynamics can be predicted with a simpler classical analogue. However, the full quantum description of our protocol allows one to retrieve the connections between the quantum living units, linked through entanglement, which is a useful complement that is absent in the classical analogue. In other words, the extra free parameters available in the superposition and entanglement of quantum states are used for describing questions regarding the collective dynamics of individuals, and this is precisely the new source of complex behavior our algorithm is able to create. In this sense, the complexity of our quantum algorithm may only be reached by a larger quantum computer, currently being built in academic institutions and companies. This might yield unexpectedly interesting outcomes but, at the same time, will increase the sensitivity to decoherence during the self-replication process.

This experimental realization of the proposed quantum algorithm represents the consolidation of the theoretical framework of quantum artificial life. The improvement in scalable quantum computers will soon allow us for more accurate quantum emulations with growing complexity towards quantum supremacy, even considering spatial variables for the individuals and a mechanism for tracing out death living units. These future developments should lead towards an autonomous character of the set of individuals, i.e., the evolution will be an intrinsic property of the system, and the desired behavior will emerge without following the instructions of a previously designed quantum algorithm. In this context, the system would be transformed into an intelligent source of quantum complexity whose evolutionary plot for a large number of individuals may not be predicted classically and, consequently, has the capacity to produce unexpected results when scaled up. An interesting question to address is to establish the relation existing between the parameters defining the fundamental processes of the model, and the emergent multiqubit quantum state. Along these lines, and following the frame of artificial life oriented genetic algorithms~\cite{alga}, we speculate about the idea of channeling this complexity to encode optimization problems by tuning the self-replication, mutation and dissipation rates that define the evolution. Furthermore, recent advances in quantum machine learning constitute a promising material to work with in the study of algorithms combining the properties of both fields, pursuing the design of intelligent and replicating quantum agents. Therefore, the creation of these quantum living units and their possible applications are expected to have deep implications in the community of quantum simulation and quantum computing in a variety of quantum platforms.

All in all, the experiments presented here entail the validation of quantum artificial life in the lab and, in particular, in cloud quantum computers as that of IBM. Still another interesting step would be the development of autonomous quantum devices following the theoretical and experimental results in quantum cellular automata \cite{qca1,qca2,qca2Bis,qca2Bis2,qca2Bis3}. Our quantum individuals are driven by an adaptation effort along the lines of a quantum Darwinian evolution, which effectively transfer the quantum information through generations of larger multiqubit entangled states. We believe that the presented results and vision, both in theory and experiments, should hoist this innovative research line as one of the leading banners in the future of quantum technologies.

\section*{Acknowledgements}
We thank Armando P\'erez-Leija and Alexander Szameit for enthusiastic and enlightening discussions. We acknowledge the use of IBM Quantum Experience for this work. The views expressed are those of the authors and do not reflect the official policy or position of IBM or the IBM Quantum Experience team. We acknowledge support from Spanish MINECO/FEDER FIS2015-69983-P, UPV/EHU new PhD program, Basque Government Programa Posdoctoral de Perfeccionamiento de Personal Investigador Doctor, Basque Government IT986-16, and Ram\'on y Cajal Grant RYC-2012-11391.

\section*{Author Contributions}
U. A.-R. performed the experiments on the IBM Quantum Experience. U. A.-R., M. S., L. L., and E. S. developed the model and analyzed the data. All authors contributed to writing the manuscript.

\section*{Additional Information}
{\bf Competing interests:} The authors declare no competing interests.

\end{document}